\def\lesssim{\mathrel{\hbox{\rlap{\hbox{\lower4pt\hbox{$\sim$}}}\hbox{$<$}}}}
\def\gtrsim{\mathrel{\hbox{\rlap{\hbox{\lower4pt\hbox{$\sim$}}}\hbox{$>$}}}}

\def\teff{$T_{\rm eff}$~}

\def\ll_lsun{log$({L/\rm L_{\odot}})$~}
\def\masa_msun{$M/ \rm M_{\odot}$~}
\def\m_mstar{$M/M_{*}$~}

\documentclass{aa}

\usepackage{graphicx}
        
\begin{document}

\title{The age and colors of massive white dwarf stars}

\author{L. G. Althaus$^{1,5}$\thanks{Member of the Carrera del Investigador
        Cient\'{\i}fico y Tecnol\'ogico, CONICET, Argentina.},
        E. Garc\'{\i}a--Berro$^{2,3}$, 
        J. Isern$^{3,4}$,  
        A. H. C\'orsico$^{1,5\star}$, \and
        R. D. Rohrmann$^{6\star}$}
\offprints{L. G. Althaus }

\institute{Facultad de Ciencias  Astron\'omicas y Geof\'{\i}sicas,
           Universidad  Nacional de  La Plata,  
           Paseo del  Bosque s/n,
           (1900) La Plata, Argentina.\
           \and
           Departament de F\'\i sica Aplicada, 
           Universitat  Polit\`ecnica de Catalunya,  
           Av. del Canal  Ol\'\i  mpic,   s/n,  
           08860  Castelldefels,  Spain\
           \and
           Institut d'Estudis  Espacials de Catalunya,  
           Ed. Nexus, c/Gran Capit\`a  2, 
           08034 Barcelona,  Spain.\
           \and
           Institut de Ci\`encies de l'Espai, CSIC,  
           Campus UAB, Facultat de Ci\`encies, Torre C--5,   
           08193   Bellaterra,   Spain\
           \and
           Instituto de Astrof\'{\i}sica La Plata, IALP, CONICET\
           \and
           Observatorio  Astron\'omico,  
           Universidad Nacional de C\'ordoba, 
           Laprida 854, 
           (5000) C\'ordoba, Argentina\\
\email{althaus@fcaglp.unlp.edu.ar, garcia@fa.upc.edu, isern@ieec.fcr.es,\\
       acorsico@fcaglp.unlp.edu.ar, rohr@oac.uncor.edu}}
\date{\today}

\abstract{}
         {We present evolutionary  calculations and colors for massive
         white dwarfs with oxygen--neon  cores for masses between 1.06
         and $1.28\,M_{\sun}$.  The evolutionary stages computed cover
         the luminosity range  from $\log(L/L_{\sun})\approx$ 0.5 down
         to $-5.2$.}  {Our cooling sequences are based on evolutionary
         calculations that take  into account the chemical composition
         expected  from  massive white  dwarf  progenitors that  burned
         carbon  in  partially  degenerate  conditions.   The  use  of
         detailed non-gray model  atmospheres provides us with accurate
         outer  boundary conditions  for  our evolving  models at  low
         effective temperatures.}  {We examine the cooling age, colors
         and magnitudes of our  sequences.  We find that massive white
         dwarfs are characterized by very short ages to such an extent
         that they reach the turn--off in their colors and become blue
         at ages well below 10 Gyr.  Extensive tabulations for massive
         white  dwarfs,  accessible  from   our  web  site,  are  also
         presented.}  {}
\authorrunning{Althaus et al.}
\titlerunning{The age and colors of massive white dwarf stars}

\maketitle


\section{Introduction}

White dwarf  stars constitute the most common  end--product of stellar
evolution.   The  present population  of  white  dwarfs thus contains  
precise information  about the  star formation rate  of our  Galaxy, as
well as about  its age, which is information that can  be accessible from their
mass and luminosity distributions,  as long as evolutionary models for
the progenitor of white dwarfs and for the white dwarfs themselves are
available.  However, most  of the information about the  birth and the
evolution of the galactic disk is concentrated at the faint end of the
white   dwarf  luminosity   function,  which   is  dominated   by  the
contribution of massive white dwarfs (D\'\i az--Pinto et al. 1994). It
is also worth mentioning at this point that the MACHO collaboration in
their first  season reported a  microlensing event with a  duration of
110 days towards  the galactic bulge (Alcock et  al.  1995).  For this
particular event  a parallax could be  obtained from the  shape of the
light curve,  from which  a mass of  $1.3^{+1.3}_{-0.6}\,M_{\sun}$ was 
then derived, indicating  that the gravitational  lens could possibly  be a
massive  oxygen--neon  (ONe)  white  dwarf  or a  neutron  star.   

The
interest in very low--luminosity white dwarfs has also increased since
the MACHO team  proposed that the microlensing events  towards the LMC
could be due  to a population of faint white  dwarfs --- see, however,
Isern et al.  (1998a), Torres  et al.  (2002), and Garc\'\i a--Berro et
al.   (2004).  Since ONe  white dwarfs  cool faster  than the  bulk of
carbon--oxygen  (CO) white  dwarfs,  it is  reasonable  to expect  that
perhaps some of the events could be due to these elusive massive white
dwarfs.   Moreover, studies about  the distribution  of masses  of the
white dwarf  population (Finley  et al.  1997;  Liebert et  al.  2005)
show the existence of a narrow sharp peak near $0.6\,M_{\sun}$, with a
tail extending  towards higher masses, with several  white dwarfs with
spectroscopically  determined  masses  within the  interval  comprised
between $1.0$ and $1.2\,M_{\sun}$.

On the other hand, theoretical evidence suggests that high--mass white
dwarfs should  have cores  composed mainly of  oxygen and neon  --- at
least  for non--rotating  stars (Dom\'\i  nguez et  al.  1996)  --- in
contrast  to average--mass  white dwarfs,  for  which carbon--oxygen
cores are  expected.  The existence  of such massive white  dwarfs has
been suggested as the result of either  binary evolution (Marsh et
al.  1997) --- see also Guerrero et al. (2004) --- or of the evolution
of  heavy--weight  intermediate--mass  single  stars (Ritossa  et  al.
1996; Garc\'\i a--Berro et al.  1997b; Iben et al. 1997, Ritossa et al.
1999).  In  particular, Garc\'{\i}a--Berro  et al.  (1997b)  found that,
when the core mass of a $9 \, M_{\sun}$ white dwarf progenitor exceeds
$\approx   1.05  \,M_{\sun}$,   carbon  is   ignited   off--center  in
semidegenerate conditions before  reaching the thermally pulsing phase
at the AGB tip.  As a result of repeated carbon--burning shell flashes
that ultimately  gave rise to  carbon exhaustion, these  authors found
that  at  the  end  of  carbon  burning  the  star  was  left  with  an
oxygen--neon  core  almost  devoid  of carbon.   After  considerable
mass--loss episodes, the progenitor remnant is expected to evolve into
the central  star of  a planetary nebula  and ultimately into  a white
dwarf with an oxygen--neon core.  A possible observational counterpart
of these ultramassive  white dwarfs would be the  single massive white
dwarf LHS~4033,  which has a mass  of $\sim 1.32\,  M_{\sun}$ (Dahn et
al.   2004).  Other  possible massive  white dwarfs  with oxygen--neon
cores would  be the magnetic  white dwarf PG~1658+441 (Schmidt  et al.
1992; Dupuis et al.  2003) --- with a mass of $\simeq 1.31\, M_{\sun}$
--- the  highly magnetic  white dwarf  RE~J0317--853 (Ferrario  et al.
1997),  which  has   a  mass  of  $\sim  1.35\,   M_{\sun}$,  and  the
ultramassive white dwarf GD~50 (Dobbie et al. 2006).

The mass--radius  relation and  the pulsational properties  of massive
white  dwarfs have  been the  subject of  recent theoretical  work: 
Althaus   et    al.    (2005)   and   C\'orsico    et   al.    (2004),
respectively. However, the evolution (namely, the cooling ages and 
colors) of  massive white dwarfs  considering the core  composition as
predicted by  the evolution  of massive progenitor  stars has  not yet
been  studied in  detail, in  sharp  contrast with  the situation  for
standard CO white dwarfs for  which very accurate cooling sequences do
exist; see,  for instance,  Salaris et  al. (2000)  and references
therein.  A first attempt to  describe the cooling of ONe white dwarfs
was performed by  Garc\'\i a--Berro et al. (1997a)  using a simplified
cooling code, but, although the equation of state employed in this work
was very detailed, the evolutionary calculations were rather simplistic
and the  adopted chemical  profiles were a  flat ONe  mixture, without
taking into  account the CO  buffer on top  of the ONe core  that full
evolutionary calculations  predict. This  paper is aimed  at precisely
filling  this  gap by  presenting  new  evolutionary calculations  for
massive white dwarfs with oxygen--neon  cores down to very low surface
luminosities  and  effective temperatures.   In  addition, we  present
colors and magnitudes for these  stars on the basis of non--gray model
atmospheres.  Detailed model atmospheres also provide us with accurate
outer  boundary conditions  for  our evolving  models.   Our paper  is
organized as follows. In \S2 we  present our input physics.  In \S3 we
discuss our  evolutionary sequences.  Finally, in the  last section we
summarize our findings and draw our conclusions.

\section{Input physics}

The evolutionary code and the starting white dwarf configurations used
in this work are essentially those used recently in the calculation of
mass--radius  relations for  massive white  dwarfs by  Althaus  et al.
(2005), and  we refer  the reader to  that paper for  further details.
Because  we are now  interested in  providing accurate  cooling times,
some major  improvements to  the input physics  assumed in  Althaus et
al.  (2005) have been  made. First,  we have  included the  release of
latent   heat   upon   crystallization.    Despite   the   fact   that
crystallization in  massive white dwarfs  takes place at higher
stellar  luminosities than do standard  carbon--oxygen
white  dwarfs,  its  impact  on  the cooling  times  is  not  entirely
negligible.  In our calculations, crystallization sets in when the ion
coupling  constant  $\Gamma$=  180,  where  $\Gamma  \equiv  Z^2  e^2
/\overline{r} k_{\rm  B} T$  and $\overline{r}$ is  the radius  of the
Wigner--Seitz sphere.  We considered  a latent heat release of $k_{\rm
B}T$ per ion, which was spread over  the range $ 175 < \Gamma < 185 $.
Phase  separation upon  crystallization  has been  shown to  introduce
negligible time delays for  massive white dwarfs (Garc\'\i a--Berro et
al.   1997a)  and, consequently,  was  not  taken into  account.
Second, for the high--density regime, we adopted the equation of state
described  in Segretain  et al.   (1994)  to account  for all  the
important contributions for  both the liquid and the  solid phases. In
particular,  this equation of  state, besides  the  contribution from
partially degenerate electrons, accounts for the exchange contribution
(Stringfellow et  al. 1990) and the  quantum (diffraction) corrections
(Hansen  \&  Vieillefosse  1975)  in  the fluid  phase,  as well as 
the  harmonic
contribution    (Chabrier   1993),   the    anaharmonic   contribution
(Stringfellow  et  al.   1990),  and  the  electrostatic  contribution
(Stringfellow et al. 1990) in  the solid phase.  For the electrons the
exchange  effects  at  finite   temperature  (Kovetz  et  al.   1972),
the correlation   contribution  (Nozi\`eres   \&  Pines   1958),  and  the
electron--ion coupling contribution  (Yakovlev \& Shalibkov 1989) were
also  taken into  account.  For  the  low--density regime,  we used  an
updated  version of  the  equation  of state  of  Magni \&  Mazzitelli
(1979).   Neutrino   emission  rates  for  pair,   photo,  plasma,  and
bremsstrahlung  processes and for high-density conductive  opacities were
taken  from Itoh  et  al.  (1994, 1996a, 1996b).
Convection was treated in the  framework of the mixing length theory as
given by  the ML2 parameterization (Tassoul et  al.  1990).  Radiative
opacities were those of OPAL (Iglesias \& Rogers 1996), complemented at
low  temperatures  with the  Alexander  \&  Ferguson (1994)  molecular
opacities.

The initial white dwarf models from which we started our calculations of
the  cooling sequences  correspond to  hot white  dwarf configurations
with  a chemical  stratification appropriate  to massive  white dwarfs
resulting from progenitor stars with solar metallicity that are expected
to  have  burned  carbon  in  semidegenerate conditions  ---  see,  for
instance  Ritossa et  al. (1996).   According to  the  calculations of
these authors, the  core is mainly composed of  $^{16}$O and $^{20}$Ne,
plus  some traces  of $^{12}$C,  $^{23}$Na,  and $^{24}$Mg  and is  the
result of  repeated shell  flashes that take  place during  the carbon--burning phase in  massive intermediate--mass stars (Garc\'{\i}a--Berro
et al.  1997b; Gil--Pons et al.  2003).  The core chemical profiles of
our models are flat throughout the core. Such profiles are expected if
Rayleigh--Taylor instabilities act to smooth--out regions with negative
molecular weight  gradients.  The outer  layer chemical stratification
consists of  a pure hydrogen envelope  of $10^{-6} M_*$  (plus a small
inner tail)  overlying a helium--dominated shell of  $4 \times 10^{-4}
M_* $  and, below that, a  buffer rich in $^{12}$C  and $^{16}$O.  The
amount of hydrogen we adopted is  an upper limit as imposed by nuclear
reactions.   All the  evolutionary sequences  considered in  this work
have the same core composition  and shell profile, which remains fixed
during the evolutionary sequences  and corresponds to that illustrated
in Fig. 4 of C\'orsico et  al.  (2004). The shape of the outer layer
chemical  profile   is  given   by  element  diffusion   at  low
luminosities.  Nevertheless,  diffusion was switched off  in the
present calculations.  Although minor changes in  the chemical profile
are expected because of the different masses of progenitor objects, we
believe that these had a negligible influence on the cooling times of our
sequences.

We  computed the  evolution of our  models in  a self--consistent
way, using  detailed non--gray model atmospheres, which also allow us
to derive  color indices  and magnitudes of  these white  dwarfs.  Our
model  atmospheres are  based  on  an ideal  equation  of state.   The
following species have been considered: H, H$_2$, e$^-$, H$^-$, H$^+$,
H$_2$$^+$, and H$_3$$^+$.  All the relevant bound--free, free--free, and
scattering processes contributing to opacity were included in our
calculations.   At  low  \teff values,  collision--induced  absorption
(CIA) from molecular hydrogen  due to collisions with H$_2$ represents
a major source  of opacity in the infrared and  dominates the shape of
the emergent  spectrum.  Collision--induced--absorption cross sections
are from  Borysow et  al.  (1997). Broadband  color indices  have been
calculated using  the optical $BVRI$  and infrared $JHK$  passbands of
Bessell  (1990)  and  Bessell  \&  Brett  (1988),  respectively,  with
calibration  constants from  Bergeron et  al.  (1997).   More specific
details about the input physics  and computational issues of the model
atmospheres  used in  this work  are described  at length  in Rohrmann
(2001).  However,  for the  purpose of this  paper it is  important to
realize that in order to compute the outer boundary conditions for our
evolving models, the values of the pressure and the temperature at the
bottom of  the atmosphere are required.  In  the calculations reported
here  non--gray model atmopheres  are fully  integrated each  time the
outer boundary conditions must be evaluated as evolution proceeds.

\begin{figure}[t]
\centering
\includegraphics[clip,width=250pt]{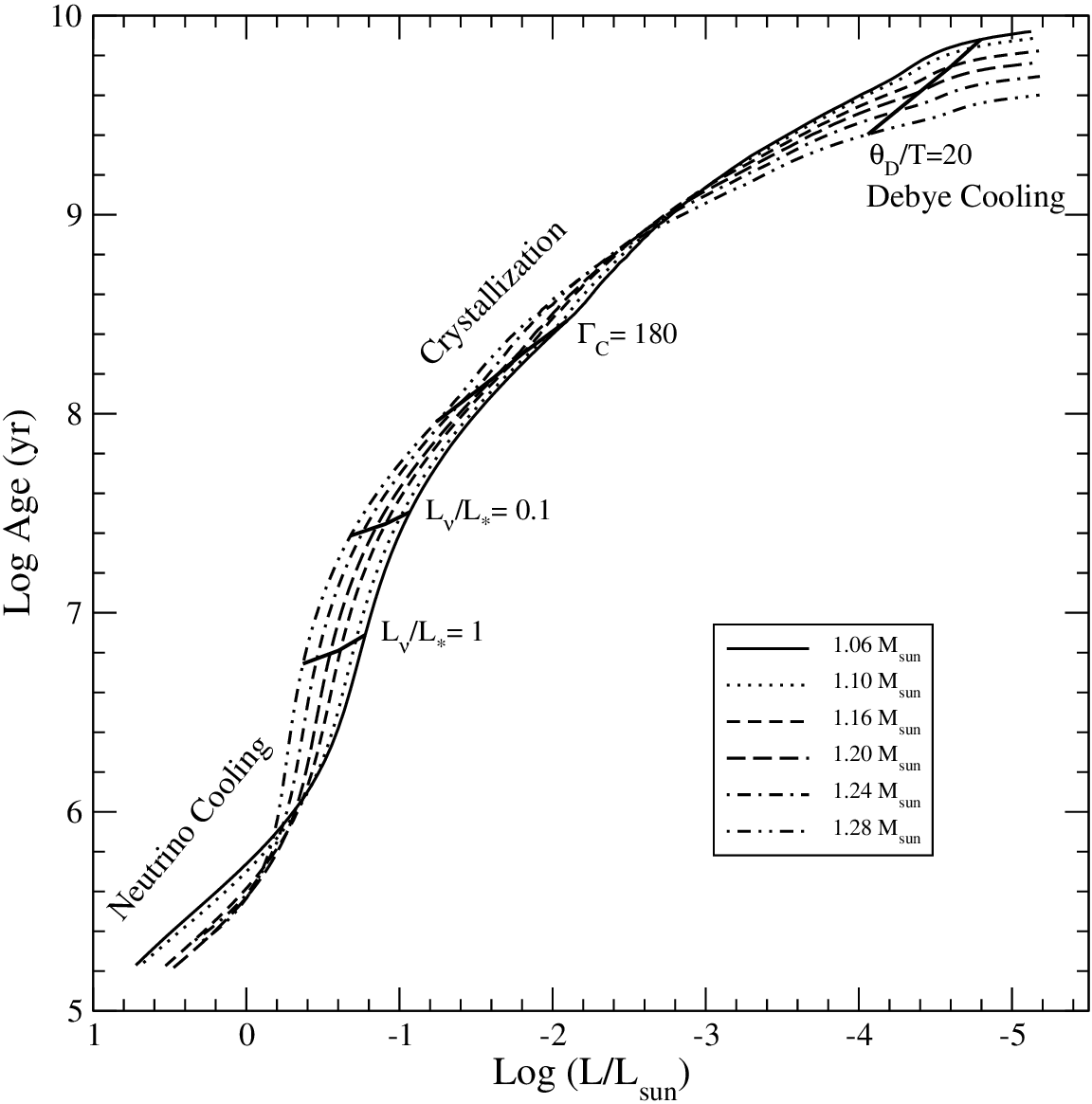}
\caption{Evolutionary times as a function of the surface luminosity for
         our massive  white dwarf models with pure  H atmospheres. The
         stages for which neutrino emission, crystallization and Debye
         cooling drive  the evolution  are indicated. The  locii along
         the  curves where  the neutrino  contribution to  the surface
         luminosity (100 and 10\%),  the onset of core crystallization
         (at   $\Gamma=  180$)  and   when  the   dimensionless  Debye
         temperature at  the core  reaches $\theta_{\rm D}/T=  20$ are
         clearly   indicated.    Note   the  relatively   short   ages
         characterizing these models  at advanced stages of evolution,
         particularly for the most massive ones.}
\label{edad_h}
\end{figure}

The   initial  stellar   models  were   derived  from   an  artificial
evolutionary procedure  starting from the full evolution  of a $0.95\,
M_{\sun}$  white  dwarf model; see  Althaus  et al.   (2005)  and
references cited  therein.  Despite the correctness  of our artificial
procedure to generate starting  white dwarf configurations, model ages
corresponding to the very first computed stages of evolution should be
taken with  some care.  However, for  ages over  10$^5$ yr, the
evolutionary  calculations  presented  here  are  independent  of  the
initial conditions.   We computed the evolution  of massive white
dwarf models with  stellar masses of 1.06, 1.10,  1.16, 1.20, 1.24, and
$1.28\, M_{\sun}$.  The lower limit  of this mass range corresponds to
the approximate  minimum mass for a  white dwarf to have  an ONe core.
The precise value  of this mass is still not  well known. White dwarfs
with masses higher than $\sim  1.1\, M_{\sun}$ have  progenitors that
proceed  through  carbon burning  (Garc\'\i  a--Berro  et al.   1997b), 
whereas white  dwarfs with masses  lower than $\sim  1.0\, M_{\sun}$
have progenitors that never ignite  carbon (Salaris et al.  1997). The
evolutionary   stages  computed  cover   the  luminosity   range  from
$\log(L/L_{\sun}) \approx 0.5$ down to $-5.2$.

In  addition  to the  calculation  of  the  evolution of  white  dwarf
sequences with pure hydrogen envelopes, we computed the evolution
of the same sequences mentioned above  but for the case of pure helium
envelopes.  In this case, we assumed a grey atmosphere.  Although
the employment of  a gray atmosphere in our  helium envelope sequences
prevents a precise  quantitative comparison with the  case of
hydrogen  envelope  sequences,  this   still  gives  us  a  reasonable
assessment of the cooling times for these helium envelope sequences.

\section{Evolutionary results}

\begin{figure}[t]
\centering
\includegraphics[clip,width=250pt]{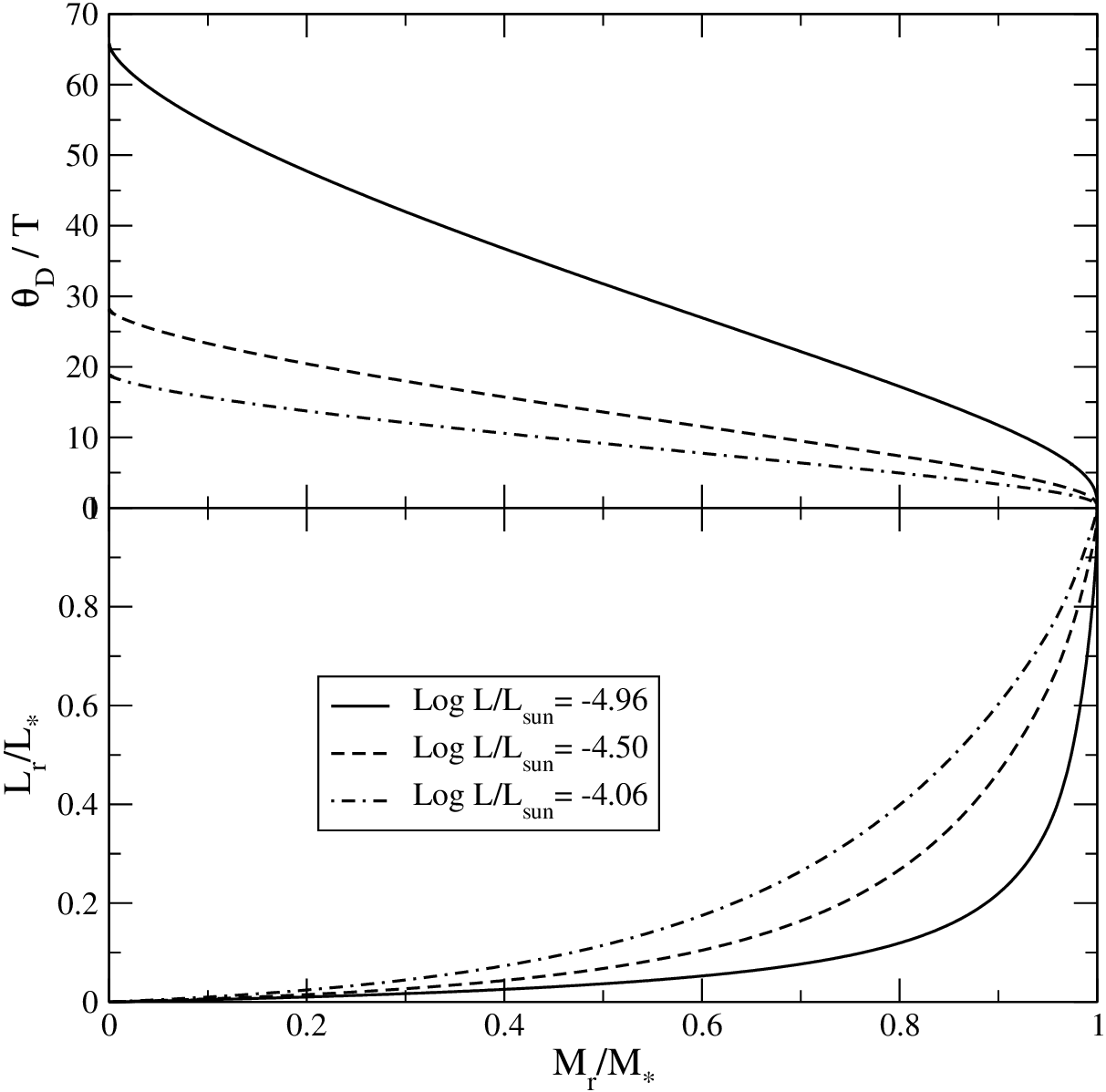}
\caption{Upper panel: The dimensionless Debye  temperature in terms of 
         the mass fraction  for three selected low--luminosity $1.28\,
         M_{\sun}$ models.   Bottom panel: Internal  luminosity versus
         the  mass fraction  for the  same models  shown in  the upper
         panel.   Note  that  most  of  the structure  of  the  lowest
         luminosity model is well inside the Debye regime.}
\label{lumis}
\end{figure}

We begin by examining  Fig.~\ref{edad_h}, which shows the evolutionary
times  as a function  of the  surface luminosity  for the  1.06, 1.10,
1.16, 1.20,  1.24, and  $1.28\, M_{\sun}$ ONe  white dwarf  models with
pure H atmospheres.  The figure  also shows (i) the locii along the curves
where the  neutrino emission  luminosity becomes 100  and 10\%  of the
surface luminosity, (ii) the onset  of core crystallization  (at $\Gamma=$
180), and (iii) the dimensionless  Debye temperature at the core when it 
reaches
$\theta_{\rm D}/T$=  20, where  the Debye temperature  is $\theta_{\rm
D}=3.48\times  10^3\langle  Z/A\rangle\rho^{1/2}$.   Several  physical
processes that  take place along the evolution  leave their signatures
in  the cooling  curve.  During  the  first stages  of evolution,  the
interior temperature is relatively high (between $T\sim 1.2$ and $1.5
\times 10^8$ K, depending on the mass of the white dwarf) and neutrino
emission  constitutes  a powerful  sink  of  energy that  considerably
affects both  the cooling timescales  of massive white dwarfs  and, as
shown in  Althaus et  al.  (2005), also  the mass--radius  relation of
these  stars.  Indeed,  neutrino emission  luminosity far  exceeds the
photon  luminosity during  the hot  white dwarf  stages.   As neutrino
emission gradually decreases, this accelerated cooling phase arrives at
its end and the slope of the  cooling curve changes, as can be seen in
Fig.~\ref{edad_h}. Note  that neutrino losses persist  further down to
luminosities  $\log(L/L_{\sun})\sim  -1$,  but  their effects  on  the
cooling curve are negligible at those evolutionary stages.

\begin{figure}[t]
\centering
\includegraphics[clip,width=250pt]{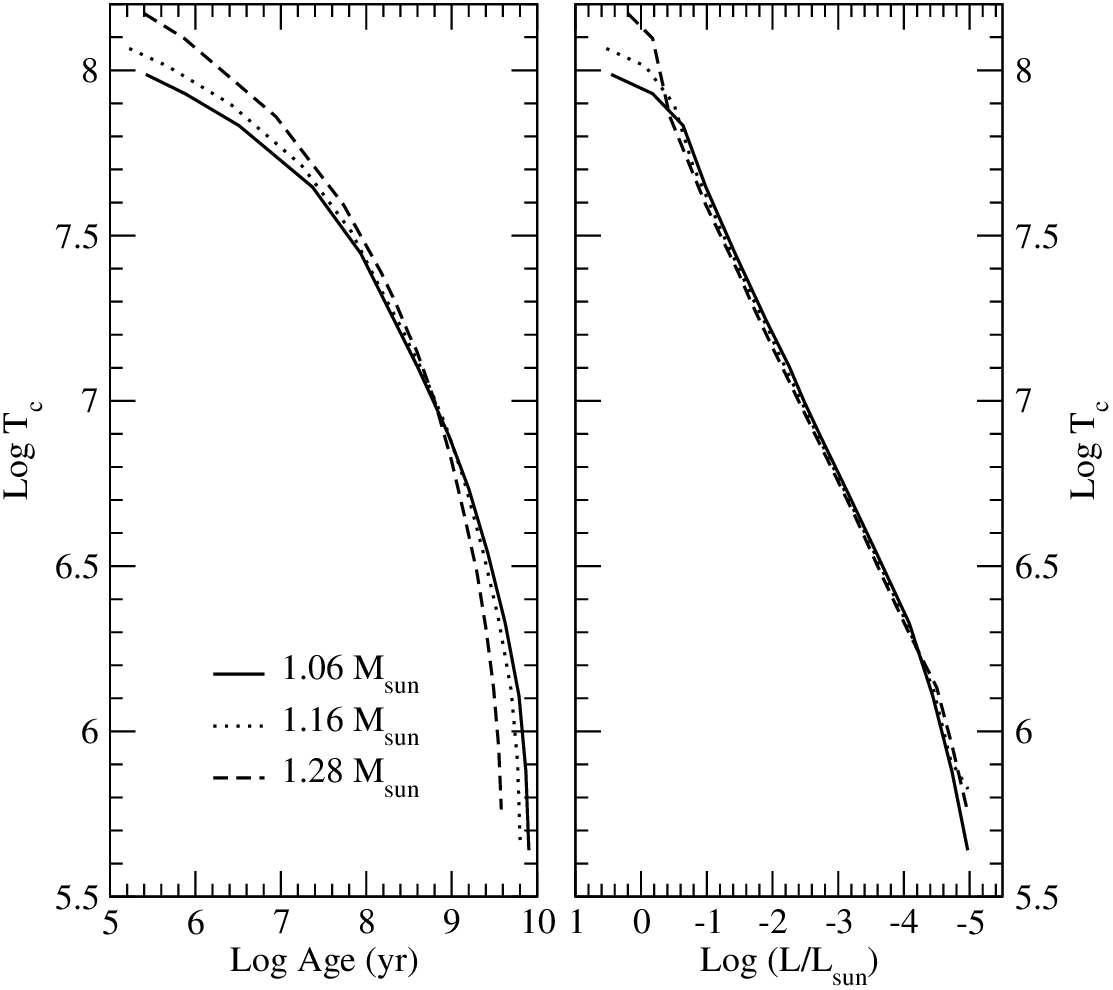}
\caption{Central  temperature versus  age and  surface luminosity  as a 
         function of the age  (left and right panels, respectively) for
         the 1.06, 1.16, and $1.28\, M_{\sun}$ sequences.}
\label{tc}
\end{figure}

The  release of  latent heat  upon  crystallization is known  to
influence the evolutionary times of  white dwarfs as well. However, in
the case of massive ONe  white dwarfs, the crystallization of the core
is  a process  that  occurs  at relatively  high  luminosities ---
$\log(L/L_{\sun})=   -2.15$    and   $-1.30$   for    the   1.06   and
$1.28\,M_{\sun}$     cooling      sequences,     respectively     (see
Fig.~\ref{edad_h}) ---  and its  impact on the  evolution of  the star
 therefore remains  moderate.   It  is thus important  to  note  that
crystallization  profoundly influences the cooling behavior of only the
less massive sequences.

\begin{figure}[t]
\centering
\includegraphics[clip,width=250pt]{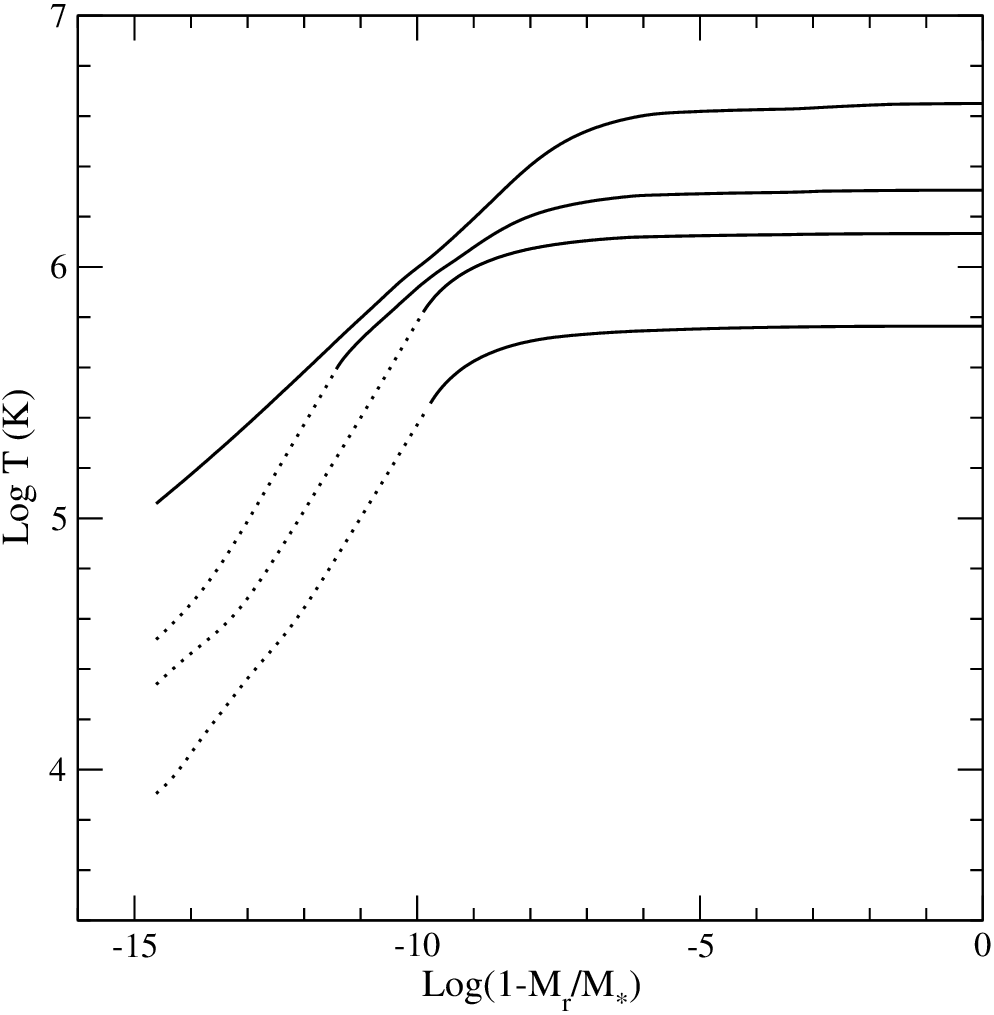}
\caption{The run  of temperature in  terms of the  outer mass fraction
         for   the    $1.28\,   M_{\sun}$   model    at   luminosities
         $\log(L/L_{\sun})= -3.25, -4.05, -4.50$, and $-4.96$ (from top
         to  bottom).  Dotted  lines  represent convectively  unstable
         layers.}
\label{tmr}
\end{figure}

The short  ages that characterize our  white dwarf models  at very low
surface  luminosities is  a noteworthy  feature. This  is particularly
noticeable  for our  most massive  models. For  instance,  the $1.28\,
M_{\sun}$ cooling sequence  reaches a luminosity of $\log(L/L_{\sun})=
-5.15$  in only 4~Gyr.   In fact  our results  indicate that  the most
massive white dwarfs could be, depending on the distance, unobservable
at ages  well below 10~Gyr with the  current observational facilities.
For  instance, if  there are  such massive  cool white  dwarfs  at the
distance  of say  the Hyades,  the  apparent magnitude  would only  be
$m_V\sim  21.5$  or  so, and  at  1  kpc,  $m_V\sim 28$,  whereas  the
$K$--band  apparent magnitudes would  be 1  or 2  magnitudes brighter,
according to our  models.  At these distances such  massive cool white
dwarfs might be observable with  present or near future facilities but
the  observations would be  very challenging.  At the  lowest computed
luminosities,   our  massive   white  dwarf   models   experience  the
development of the  so--called fast Debye cooling.  In  this sense, we
would like  to mention at  this point of  the discussion that,  for the
$1.28\,M_{\sun}$ model at $\log(L/L_{\sun})= -4.95$, the dimensionless
Debye  temperature remains well  above 10  for about  the 95\%  of the
structure (that is, most of the star is within the Debye regime), with
the consequence that the thermal  content goes rapidly to zero in this
region.   For  less massive  CO  white  dwarfs,  this takes  place  at
luminosities of about  $\log(L/L_{\sun})=  -6$ (D'Antona  \&  Mazzitelli
1989).  This fact is  reflected in Fig.~\ref{lumis}, which displays the
run of  the Debye temperature  and the fractional luminosity  for some
selected   low--luminosity  stages   of  the   $1.28\,M_{\sun}$  model
sequence. Note that  for the coolest models, the  luminosity output is
almost negligible  in the inner regions  of the star and  that only in
the outermost  layers are the ratio $\theta_{\rm D}/T$  and the degeneracy
parameter not  very  large.   Hence, these  layers  can still be
compressionally heated, making the  main energy source
of  the  star.  In  fact,  the  compression  of the  outermost  layers
provides the bulk of the star  luminosity; see Isern et al. (1998b)
for a detailed description of the cooling of white dwarfs.

For a better understanding of  the physical processes occurring in the
interior of  our massive white  dwarf models we show  in Fig.~\ref{tc}
the  run   of  the  central  temperatures  versus   ages  and  surface
luminosities (left  and right panel, respectively) for  the 1.06, 1.16
and $1.28\,  M_{\sun}$ model  sequences. The changes  in slope  of the
$\log  T_{\rm  c}$  versus  $\log(L/L_{\sun})$ diagram  at  both  high
luminosities and  at the low  luminosity end are noticeable.   At high
luminosity, it reflects  the end of neutrino dominated  regime. At low
luminosity, the  increase in  the slope of  the curve occurs  when the
hydrogen--rich  outer   convection  zone  approaches   the  isothermal
degenerate core  (see Fig.~\ref{tmr}).  As  a result, the  white dwarf
has  additional  thermal energy  to  radiate  (D'Antona \&  Mazzitelli
1989).  This  helps to understand the lengthening  of the evolutionary
cooling   times   occurring  at   $\log(L/L_{\sun})   \approx  -4$   in
Fig.~\ref{edad_h}.

\begin{figure}[t]
\centering
\includegraphics[clip,width=250pt]{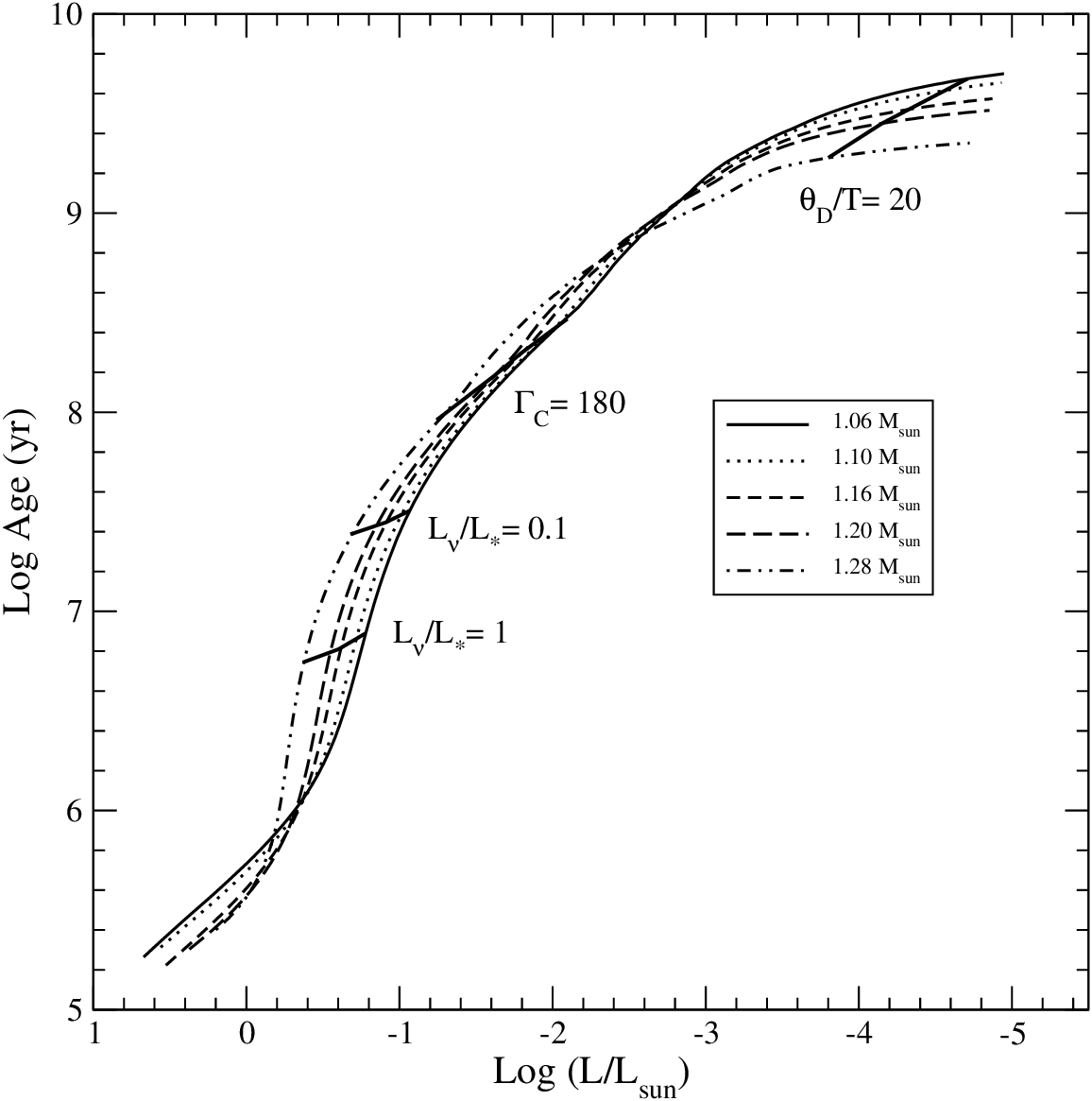}
\caption{Same as  Fig.  \ref{edad_h} but for  white dwarf  models with 
         pure He atmospheres.}
\label{edad_he}
\end{figure}

\begin{figure}[t]
\centering
\includegraphics[clip,width=250pt]{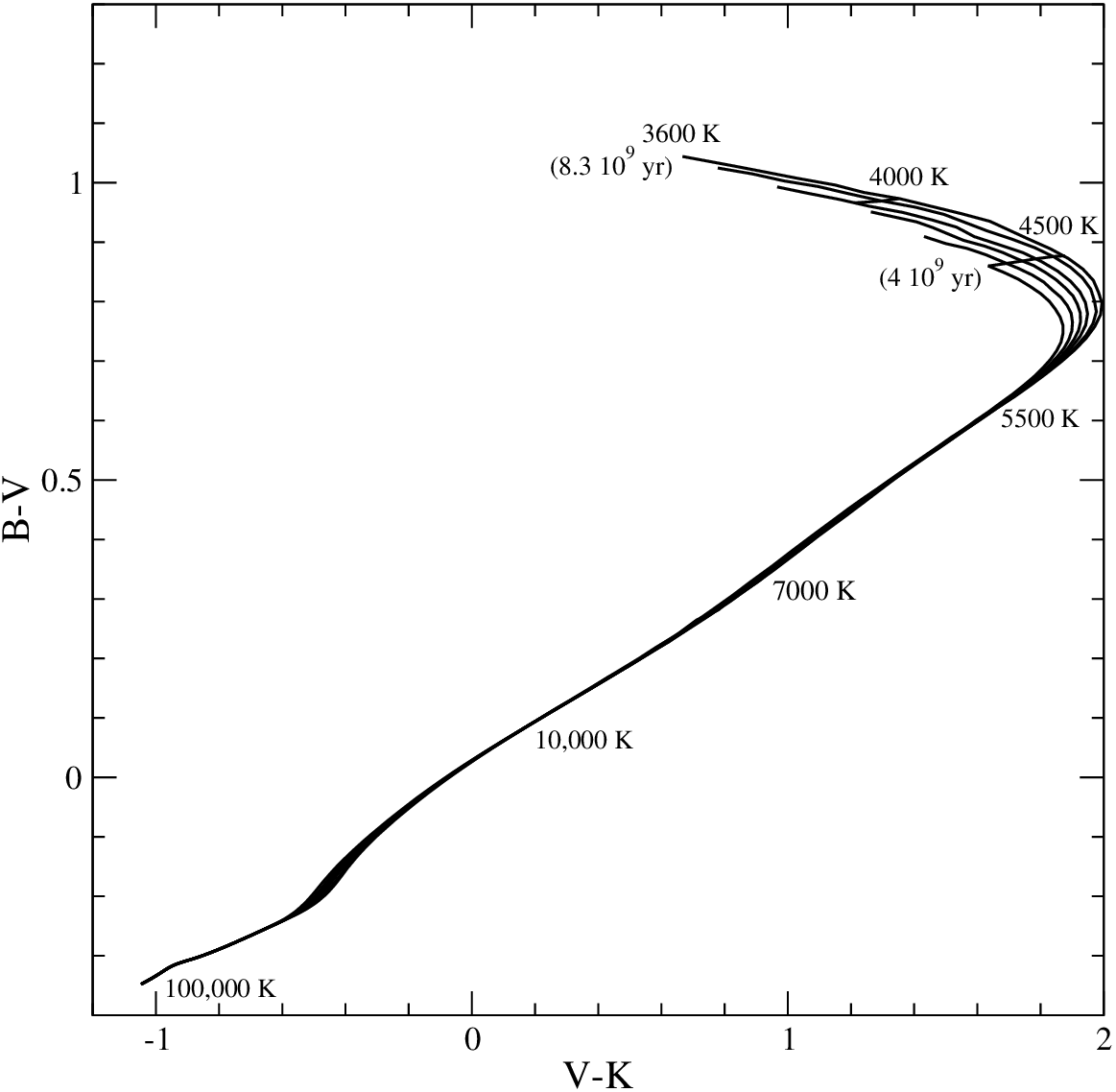}
\caption{($B-V, V-K$) color--color diagram for our massive white dwarf 
         cooling sequences  with masses  1.06, 1.10, 1.16,  1.20, 1.24,
         and $1.28 \,M_{\sun}$ (from top to bottom). Lines tracing 
         the locii  of equal effective temperatures  are labelled with
         the corresponding values. For the 1.06 and $1.28 \, M_{\sun}$
         sequences, the ages at  \teff= 3600 and 4500~K, respectively,
         are  also indicated.   Note  that all  the cooling  sequences
         exhibit  a   pronounced  turn--off  at   advanced  stages  of
         evolution.   Below   4500~K,  massive  white   dwarfs  become
         markedly blue in this  diagram within cooling times much less
         than 10~Gyr.}
\label{v_kvsb_v}
\end{figure}

\begin{figure}[t]
\centering
\includegraphics[clip,width=250pt]{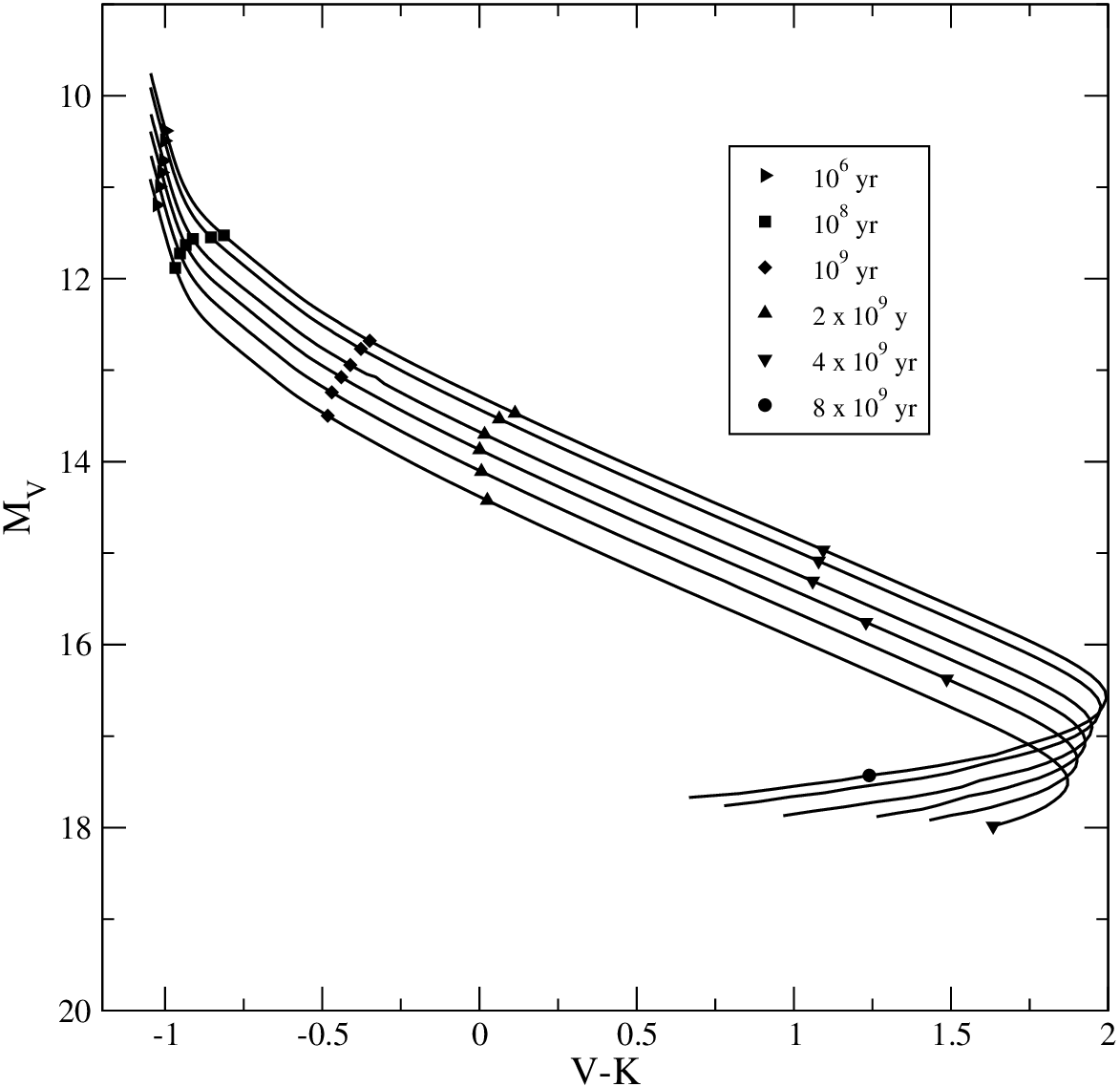}
\caption{Absolute visual magnitude $M_V$ in terms of  the color  index 
         $V-K$ for our massive white dwarf sequences with masses 1.06,
         1.10,  1.16, 1.20,  1.24  and $1.28\,M_{\sun}$  (from top  to
         bottom).   On each  curve, filled  symbols  represent cooling
         ages,  as   indicated  within  the   inset.   The  pronounced
         turn--off at  advanced stages  of the evolution  is apparent.
         Note that for ages well  within 10 Gyr, all our sequences are
         expected to become markedly blue in this diagram.}
\label{mv_vk}
\end{figure}

Non--negligible differences in the  cooling of white dwarfs arise from
the different thicknesses  of the H envelopes with  which white dwarfs
proceed during  their cooling track.   To assess such  differences, we
 considered the evolution of  massive white dwarfs for the extreme
situation of pure helium  envelopes.  The resulting evolutionary times
are  displayed  in  Fig.~\ref{edad_he}.   At advanced  stages in  the
evolution, the central  temperature of the models is  strongly tied to
the  details of the  outer layer's  chemical stratification.   This fact
starts to  manifest itself  when the boundary  of the  degenerate core
reaches the base of the  H envelope at $\log(L/L_{\sun})\sim -2.5$;
see Tassoul et al. (1990) for details.  As a result, white dwarfs with
helium envelopes (and hence more transparent) evolve more rapidly than
those  white  dwarfs  with  H  envelopes, as  is  clear  by  examining
Fig.~\ref{edad_he}. As  expected, the helium sequences  will reach the
Debye  cooling  conditions earlier  than  their hydrogen  counterparts
(compare  Figs.~\ref{edad_h} and \ref{edad_he}).   Note that  the
$1.28\,  M_{\sun}$   white  dwarf  cooling  sequence   with  a  helium
atmosphere needs  only 2.24 Gyr to reach  $\log(L/L_{\sun})= -4.70$, a
factor  about  1.5 less  than  the age  required by  the pure  H
counterpart. At ages  below 5 Gyr, most of  our cooling sequences with
pure He atmospheres will have cooled down to below $\log(L/L_{\sun})
\approx -5.5$.

\begin{figure}[t]
\centering
\includegraphics[clip,width=250pt]{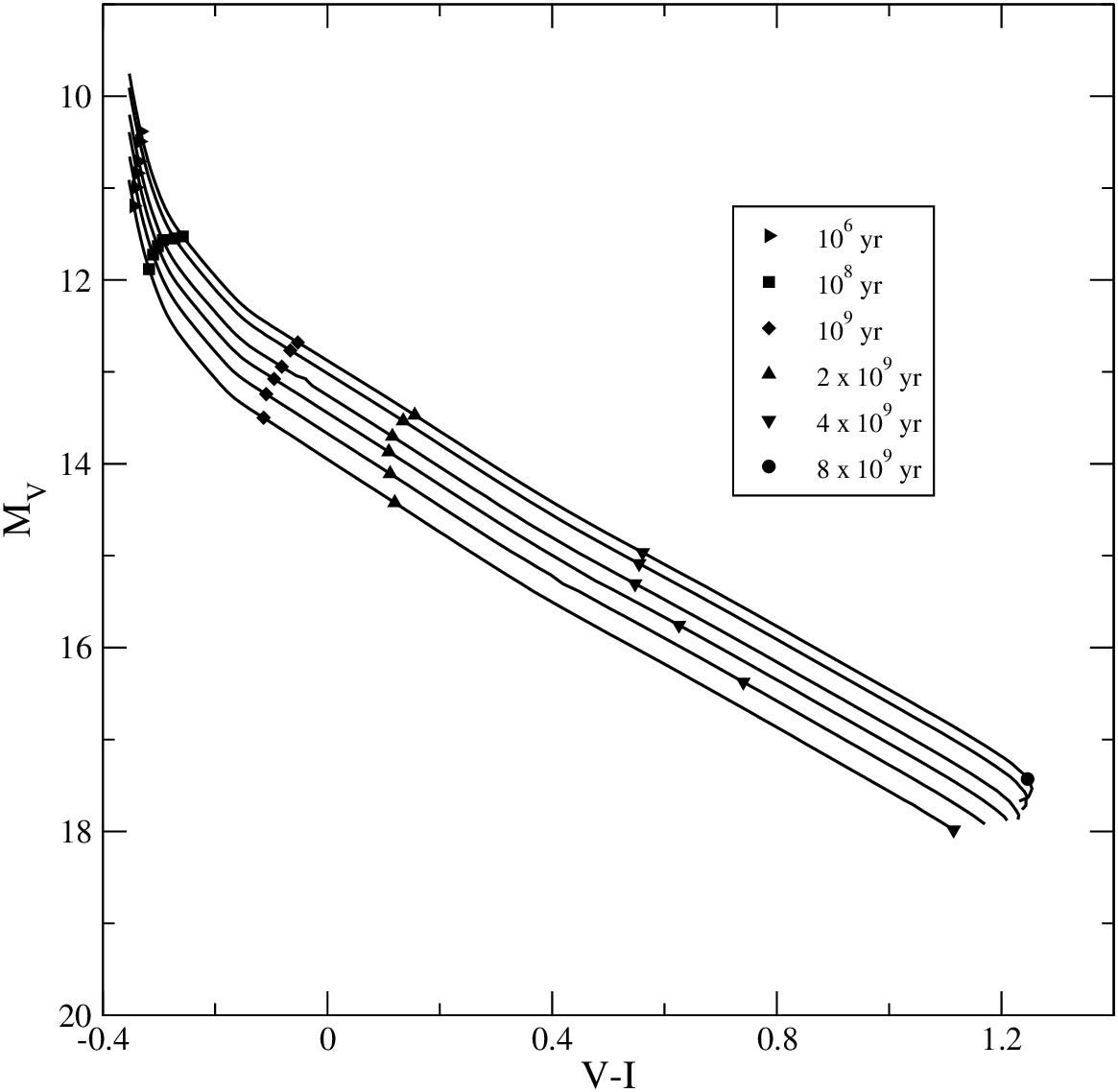}
\caption{Same as Fig.\ref{mv_vk} but for the color index $V-I$. Unlike
         the  $V-K$, this  index has  not yet  reached the
         turn--off point even for the coolest computed models.}
\label{mv_vi}
\end{figure}

The  fast  evolution  of  our  massive white  dwarfs  at  low  surface
luminosities raises the possibility  that these white dwarfs with pure
H  atmospheres reach  the turn--off  in their  colors and  become blue
afterwards  within relatively  short ages.  To  elaborate this
point further,  we show  in Fig.~\ref{v_kvsb_v}  the evolution  of our  pure H
white  dwarf  cooling  sequences  in  the  ($B-V,  V-K$)  color--color
diagram.  As  a  result  of  the  collision--induced  absorption  from
molecular  hydrogen, a  process  that reduces  the  infrared flux  and
forces  radiation to  emerge at  higher frequencies,  very  cool white
dwarfs are  expected to  become bluer as  they age (Hansen  1998).  As
shown  in  Fig.~\ref{v_kvsb_v}, this  effect  is  also  present in  our
massive white  dwarf sequences.  Indeed,  all our sequences  exhibit a
pronounced  turn--off in  their color  indices at  advanced  stages of
evolution  and become  bluer with  further evolution.  In particular,
below 4500 K, massive white dwarfs become markedly bluer in the $(B-V,
V-K)$ two--color diagram. The remarkable point is that the turn to the
blue happens  within cooling  times much shorter  than 10 Gyr.   If we
extrapolate  our results,  we  should thus  expect  dim massive  white
dwarfs characterized by relatively short ages to exhibit blue colors.

\begin{table*}
\caption{Selected stages for 1.06, 1.16, and $1.28\, M_{\sun}$ white dwarf 
         models with H atmospheres. }
\begin{tabular}{lcccccccccc}
\hline
\hline
$ M_*/M_{\sun}$ & $T_{\rm eff}$ (K) & $\log(L/L_{\sun})$& Age (10$^8$ yr)
& $U-V$ & $V-R$ &  $V-I$ & $V-K$ & $B-V$ & $J-H$ & $M_V$\\
\hline
  1.06 & 90000 &  0.55   &  0.0006 &  $-1.66$ &  $-0.15$ &  $-0.35$ &  $-1.04$ &  $-0.35$ &  $-0.15$ &   9.86 \\
  1.06 & 85000 &  0.44   &  0.0010 &  $-1.65$ &  $-0.14$ &  $-0.35$ &  $-1.03$ &  $-0.34$ &  $-0.15$ &   9.93 \\
  1.06 & 80000 &  0.33   &  0.0015 &  $-1.65$ &  $-0.14$ &  $-0.35$ &  $-1.03$ &  $-0.34$ &  $-0.15$ &   9.99 \\
  1.06 & 75000 &  0.22   &  0.0021 &  $-1.64$ &  $-0.14$ &  $-0.34$ &  $-1.02$ &  $-0.34$ &  $-0.15$ &  10.06 \\
  1.06 & 70000 &  0.09   &  0.0030 &  $-1.63$ &  $-0.14$ &  $-0.34$ &  $-1.02$ &  $-0.34$ &  $-0.15$ &  10.13 \\
  1.06 & 65000 & $-0.04$ &  0.0042 &  $-1.62$ &  $-0.14$ &  $-0.34$ &  $-1.01$ &  $-0.34$ &  $-0.15$ &  10.21 \\
  1.06 & 60000 & $-0.18$ &  0.0060 &  $-1.61$ &  $-0.14$ &  $-0.34$ &  $-1.01$ &  $-0.34$ &  $-0.14$ &  10.30 \\
  1.06 & 55000 & $-0.34$ &  0.0091 &  $-1.60$ &  $-0.14$ &  $-0.33$ &  $-1.00$ &  $-0.33$ &  $-0.14$ &  10.39 \\
  1.06 & 50000 & $-0.51$ &  0.0163 &  $-1.59$ &  $-0.14$ &  $-0.33$ &  $-0.99$ &  $-0.33$ &  $-0.14$ &  10.50 \\
  1.06 & 45000 & $-0.70$ &  0.0450 &  $-1.57$ &  $-0.13$ &  $-0.32$ &  $-0.98$ &  $-0.33$ &  $-0.14$ &  10.62 \\
  1.06 & 40000 & $-0.91$ &  0.1655 &  $-1.54$ &  $-0.13$ &  $-0.32$ &  $-0.96$ &  $-0.32$ &  $-0.14$ &  10.79 \\
  1.06 & 35000 & $-1.15$ &  0.4139 &  $-1.48$ &  $-0.13$ &  $-0.30$ &  $-0.93$ &  $-0.31$ &  $-0.14$ &  11.03 \\
  1.06 & 30000 & $-1.42$ &  0.8407 &  $-1.38$ &  $-0.12$ &  $-0.27$ &  $-0.86$ &  $-0.30$ &  $-0.12$ &  11.39 \\
  1.06 & 25000 & $-1.74$ &  1.5980 &  $-1.24$ &  $-0.10$ &  $-0.23$ &  $-0.73$ &  $-0.27$ &  $-0.10$ &  11.77 \\
  1.06 & 20000 & $-2.13$ &  3.1266 &  $-1.07$ &  $-0.08$ &  $-0.17$ &  $-0.58$ &  $-0.24$ &  $-0.08$ &  12.18 \\
  1.06 & 15000 & $-2.64$ &  8.2000 &  $-0.88$ &  $-0.03$ &  $-0.07$ &  $-0.39$ &  $-0.15$ &  $-0.05$ &  12.61 \\
  1.06 & 10000 & $-3.35$ & 20.8497 &  $-0.61$ &   0.10   &   0.18   &   0.17   &   0.09   &   0.04   &  13.57 \\
  1.06 &  9000 & $-3.53$ & 25.1306 &  $-0.54$ &   0.14   &   0.27   &   0.41   &   0.16   &   0.09   &  13.93 \\
  1.06 &  8500 & $-3.63$ & 27.8027 &  $-0.49$ &   0.17   &   0.33   &   0.54   &   0.20   &   0.12   &  14.14 \\
  1.06 &  8000 & $-3.74$ & 30.9201 &  $-0.42$ &   0.19   &   0.38   &   0.68   &   0.25   &   0.15   &  14.35 \\
  1.06 &  7500 & $-3.85$ & 34.3451 &  $-0.33$ &   0.23   &   0.45   &   0.84   &   0.30   &   0.18   &  14.59 \\
  1.06 &  7000 & $-3.97$ & 38.4651 &  $-0.20$ &   0.26   &   0.52   &   1.01   &   0.37   &   0.20   &  14.85 \\
  1.06 &  6500 & $-4.10$ & 43.1576 &  $-0.04$ &   0.31   &   0.62   &   1.21   &   0.45   &   0.24   &  15.15 \\
  1.06 &  6000 & $-4.24$ & 49.3454 &   0.15   &   0.36   &   0.72   &   1.45   &   0.54   &   0.27   &  15.50 \\
  1.06 &  5500 & $-4.39$ & 58.5783 &   0.37   &   0.42   &   0.85   &   1.73   &   0.65   &   0.32   &  15.94 \\
  1.06 &  5000 & $-4.56$ & 67.6240 &   0.62   &   0.50   &   0.99   &   1.98   &   0.76   &   0.33   &  16.44 \\
  1.06 &  4500 & $-4.74$ & 74.3322 &   0.86   &   0.57   &   1.14   &   1.87   &   0.88   &   0.13   &  16.95 \\
  1.06 &  4000 & $-4.95$ & 79.5875 &   1.07   &   0.63   &   1.24   &   1.33   &   0.98   &  $-0.15$ &  17.39 \\
  1.06 &  3600 & $-5.12$ & 83.0555 &   1.21   &   0.65   &   1.23   &   0.69   &   1.04   &  $-0.28$ &  17.66 \\
\\
  1.16 & 90000 &  0.38   &  0.0005 &  $-1.66$ &  $-0.15$ &  $-0.35$ &  $-1.04$ &  $-0.35$ &  $-0.15$ &  10.29 \\
  1.16 & 85000 &  0.27   &  0.0008 &  $-1.65$ &  $-0.14$ &  $-0.35$ &  $-1.03$ &  $-0.34$ &  $-0.15$ &  10.35 \\
  1.16 & 80000 &  0.16   &  0.0014 &  $-1.65$ &  $-0.14$ &  $-0.35$ &  $-1.03$ &  $-0.34$ &  $-0.15$ &  10.42 \\
  1.16 & 75000 &  0.05   &  0.0021 &  $-1.64$ &  $-0.14$ &  $-0.34$ &  $-1.02$ &  $-0.34$ &  $-0.15$ &  10.48 \\
  1.16 & 70000 & $-0.08$ &  0.0031 &  $-1.63$ &  $-0.14$ &  $-0.34$ &  $-1.02$ &  $-0.34$ &  $-0.15$ &  10.55 \\
  1.16 & 65000 & $-0.21$ &  0.0050 &  $-1.62$ &  $-0.14$ &  $-0.34$ &  $-1.01$ &  $-0.34$ &  $-0.15$ &  10.63 \\
  1.16 & 60000 & $-0.35$ &  0.0090 &  $-1.61$ &  $-0.14$ &  $-0.34$ &  $-1.01$ &  $-0.34$ &  $-0.14$ &  10.71 \\
  1.16 & 55000 & $-0.51$ &  0.0254 &  $-1.60$ &  $-0.14$ &  $-0.33$ &  $-1.00$ &  $-0.33$ &  $-0.14$ &  10.80 \\
  1.16 & 50000 & $-0.68$ &  0.0963 &  $-1.59$ &  $-0.14$ &  $-0.33$ &  $-0.99$ &  $-0.33$ &  $-0.14$ &  10.91 \\
  1.16 & 45000 & $-0.86$ &  0.2369 &  $-1.57$ &  $-0.13$ &  $-0.32$ &  $-0.98$ &  $-0.33$ &  $-0.14$ &  11.04 \\
  1.16 & 40000 & $-1.07$ &  0.4554 &  $-1.54$ &  $-0.13$ &  $-0.32$ &  $-0.96$ &  $-0.32$ &  $-0.14$ &  11.20 \\
  1.16 & 35000 & $-1.31$ &  0.7979 &  $-1.48$ &  $-0.13$ &  $-0.30$ &  $-0.93$ &  $-0.31$ &  $-0.14$ &  11.43 \\
  1.16 & 30000 & $-1.58$ &  1.3593 &  $-1.38$ &  $-0.12$ &  $-0.27$ &  $-0.86$ &  $-0.30$ &  $-0.12$ &  11.80 \\
  1.16 & 25000 & $-1.90$ &  2.4250 &  $-1.24$ &  $-0.10$ &  $-0.23$ &  $-0.73$ &  $-0.27$ &  $-0.10$ &  12.17 \\
  1.16 & 20000 & $-2.29$ &  5.1659 &  $-1.07$ &  $-0.08$ &  $-0.17$ &  $-0.58$ &  $-0.24$ &  $-0.08$ &  12.57 \\
  1.16 & 15000 & $-2.80$ & 10.8230 &  $-0.89$ &  $-0.03$ &  $-0.08$ &  $-0.40$ &  $-0.15$ &  $-0.05$ &  12.97 \\
  1.16 & 10000 & $-3.50$ & 22.5144 &  $-0.62$ &   0.10   &   0.18   &   0.18   &   0.09   &   0.04   &  13.97 \\
  1.16 &  9500 & $-3.59$ & 24.4627 &  $-0.59$ &   0.12   &   0.23   &   0.29   &   0.12   &   0.07   &  14.15 \\
  1.16 &  9000 & $-3.69$ & 26.6816 &  $-0.55$ &   0.14   &   0.28   &   0.41   &   0.16   &   0.09   &  14.33 \\
  1.16 &  8500 & $-3.79$ & 29.1833 &  $-0.50$ &   0.17   &   0.33   &   0.54   &   0.20   &   0.12   &  14.53 \\
  1.16 &  8000 & $-3.89$ & 32.0665 &  $-0.42$ &   0.20   &   0.38   &   0.69   &   0.25   &   0.15   &  14.75 \\
  1.16 &  7500 & $-4.01$ & 35.1530 &  $-0.33$ &   0.23   &   0.45   &   0.84   &   0.31   &   0.17   &  14.98 \\
  1.16 &  7000 & $-4.13$ & 38.4451 &  $-0.20$ &   0.26   &   0.53   &   1.01   &   0.37   &   0.20   &  15.23 \\
  1.16 &  6500 & $-4.26$ & 42.0717 &  $-0.03$ &   0.31   &   0.62   &   1.21   &   0.45   &   0.23   &  15.53 \\
  1.16 &  6000 & $-4.40$ & 47.0224 &   0.16   &   0.36   &   0.72   &   1.45   &   0.55   &   0.27   &  15.89 \\
  1.16 &  5500 & $-4.55$ & 53.6414 &   0.37   &   0.42   &   0.85   &   1.73   &   0.65   &   0.31   &  16.32 \\
  1.16 &  5000 & $-4.71$ & 58.8242 &   0.61   &   0.49   &   0.99   &   1.94   &   0.76   &   0.31   &  16.82 \\
  1.16 &  4500 & $-4.90$ & 62.5536 &   0.85   &   0.57   &   1.13   &   1.79   &   0.87   &   0.09   &  17.32 \\
  1.16 &  4000 & $-5.10$ & 65.4856 &   1.05   &   0.62   &   1.22   &   1.21   &   0.97   &  $-0.17$ &  17.74 \\
\hline
\hline
\end{tabular}
\end{table*}
\setcounter{table}{0}
\begin{table*}
\caption{Continued.}
\begin{tabular}{lcccccccccc}
\hline
\hline
$ M_*/M_{\sun}$ & $T_{\rm eff}$ (K) & $\log(L/L_{\sun})$& Age (10$^8$ yr)
& $U-V$ & $V-R$ &  $V-I$ & $V-K$ & $B-V$ & $J-H$ & $M_V$\\
\hline
  1.28 & 90000 &  0.08   &  0.0008 &  $-1.66$ &  $-0.15$ &  $-0.35$ &  $-1.04$ &  $-0.35$ &  $-0.15$ &  11.03 \\
  1.28 & 85000 & $-0.02$ &  0.0017 &  $-1.65$ &  $-0.14$ &  $-0.35$ &  $-1.03$ &  $-0.34$ &  $-0.15$ &  11.09 \\
  1.28 & 80000 & $-0.13$ &  0.0035 &  $-1.65$ &  $-0.14$ &  $-0.35$ &  $-1.03$ &  $-0.34$ &  $-0.15$ &  11.15 \\
  1.28 & 75000 & $-0.25$ &  0.0127 &  $-1.64$ &  $-0.14$ &  $-0.34$ &  $-1.02$ &  $-0.34$ &  $-0.15$ &  11.22 \\
  1.28 & 70000 & $-0.37$ &  0.0543 &  $-1.63$ &  $-0.14$ &  $-0.34$ &  $-1.02$ &  $-0.34$ &  $-0.15$ &  11.30 \\
  1.28 & 65000 & $-0.51$ &  0.1222 &  $-1.62$ &  $-0.14$ &  $-0.34$ &  $-1.01$ &  $-0.34$ &  $-0.15$ &  11.37 \\
  1.28 & 60000 & $-0.65$ &  0.2173 &  $-1.61$ &  $-0.14$ &  $-0.34$ &  $-1.01$ &  $-0.34$ &  $-0.14$ &  11.46 \\
  1.28 & 55000 & $-0.80$ &  0.3476 &  $-1.60$ &  $-0.14$ &  $-0.33$ &  $-1.00$ &  $-0.33$ &  $-0.14$ &  11.55 \\
  1.28 & 50000 & $-0.97$ &  0.5251 &  $-1.59$ &  $-0.14$ &  $-0.33$ &  $-0.99$ &  $-0.33$ &  $-0.14$ &  11.65 \\
  1.28 & 45000 & $-1.16$ &  0.7751 &  $-1.57$ &  $-0.13$ &  $-0.32$ &  $-0.98$ &  $-0.33$ &  $-0.14$ &  11.77 \\
  1.28 & 40000 & $-1.37$ &  1.1623 &  $-1.54$ &  $-0.13$ &  $-0.32$ &  $-0.96$ &  $-0.32$ &  $-0.14$ &  11.93 \\
  1.28 & 35000 & $-1.60$ &  1.8961 &  $-1.48$ &  $-0.13$ &  $-0.30$ &  $-0.93$ &  $-0.31$ &  $-0.14$ &  12.15 \\
  1.28 & 30000 & $-1.87$ &  3.0428 &  $-1.38$ &  $-0.12$ &  $-0.27$ &  $-0.86$ &  $-0.30$ &  $-0.12$ &  12.51 \\
  1.28 & 25000 & $-2.19$ &  4.8143 &  $-1.24$ &  $-0.10$ &  $-0.23$ &  $-0.73$ &  $-0.27$ &  $-0.10$ &  12.89 \\
  1.28 & 20000 & $-2.58$ &  7.7871 &  $-1.07$ &  $-0.07$ &  $-0.16$ &  $-0.58$ &  $-0.23$ &  $-0.08$ &  13.29 \\
  1.28 & 15000 & $-3.08$ & 12.2934 &  $-0.92$ &  $-0.03$ &  $-0.08$ &  $-0.41$ &  $-0.14$ &  $-0.06$ &  13.64 \\
  1.28 & 10000 & $-3.79$ & 21.7946 &  $-0.64$ &   0.10   &   0.19   &   0.19   &   0.09   &   0.05   &  14.70 \\
  1.28 &  9500 & $-3.88$ & 23.0131 &  $-0.61$ &   0.12   &   0.23   &   0.30   &   0.13   &   0.07   &  14.87 \\
  1.28 &  9000 & $-3.97$ & 24.2524 &  $-0.56$ &   0.14   &   0.28   &   0.42   &   0.16   &   0.09   &  15.05 \\
  1.28 &  8500 & $-4.07$ & 25.4978 &  $-0.50$ &   0.17   &   0.33   &   0.55   &   0.21   &   0.12   &  15.25 \\
  1.28 &  8000 & $-4.18$ & 26.8294 &  $-0.43$ &   0.20   &   0.39   &   0.69   &   0.25   &   0.15   &  15.45 \\
  1.28 &  7500 & $-4.29$ & 28.1849 &  $-0.32$ &   0.23   &   0.45   &   0.84   &   0.31   &   0.17   &  15.68 \\
  1.28 &  7000 & $-4.41$ & 29.6612 &  $-0.19$ &   0.27   &   0.53   &   1.01   &   0.38   &   0.20   &  15.94 \\
  1.28 &  6500 & $-4.54$ & 31.6081 &  $-0.03$ &   0.31   &   0.62   &   1.21   &   0.46   &   0.23   &  16.24 \\
  1.28 &  6000 & $-4.68$ & 34.4408 &   0.16   &   0.36   &   0.72   &   1.45   &   0.55   &   0.27   &  16.60 \\
  1.28 &  5500 & $-4.83$ & 36.7766 &   0.37   &   0.42   &   0.85   &   1.71   &   0.65   &   0.31   &  17.03 \\
  1.28 &  5000 & $-5.00$ & 38.4017 &   0.60   &   0.49   &   0.99   &   1.87   &   0.76   &   0.28   &  17.52 \\
  1.28 &  4600 & $-5.14$ & 39.6009 &   0.78   &   0.55   &   1.09   &   1.72   &   0.84   &   0.09   &  17.90 \\
\hline
\hline
\end{tabular}
\end{table*}

The molecular  hydrogen absorption at low  effective temperatures also
affects the evolution of our models in the color-magnitude diagram, as
 shown in Figs.~\ref{mv_vk}  and \ref{mv_vi}, which  display the
run of the  absolute visual magnitude $M_V$ in terms  of the $V-K$ and
$V-I$  color  indices,  respectively.   For  the  evolutionary  stages
computed in  this work,  the turn  to the blue  is noticeable  for the
$V-K$ color  index. Note  that in this  diagram, all  our  sequences   
are  expected  to  become  markedly  blue for relatively short ages.
Specifically, our sequences have cooling  ages between 3.7 and 6.8 Gyr
(depending on  the stellar mass  value) at the turn--off  point, which
occurs  at $M_V  \approx  17$.   For later  stages,  Debye cooling  is
dominant and evolution indeed proceeds very quickly.  Note  as well that
in  the $(M_V,V-K)$ diagram  our sequences  remain brighter  than $M_V
\approx 18$ at  advanced stages.  For the lowest  luminosities we have
driven  our cooling  tracks, our  sequences have  not yet  reached the
turn--off  point in  the $(M_V,V-I)$  diagram, as  inferred  from Fig.
~\ref{mv_vi}.  For the turn to the blue to occur, the evolution should
have  proceeded to  lower effective  temperatures than  those computed
here. Although this would certainly take place in ages shorter than 10
Gyr, the expected surface luminosity  of the models would be extremely
low to be detected --- below $\log(L/L_{\sun})= -5.5$.

\begin{figure}[t]
\centering
\includegraphics[clip,width=250pt]{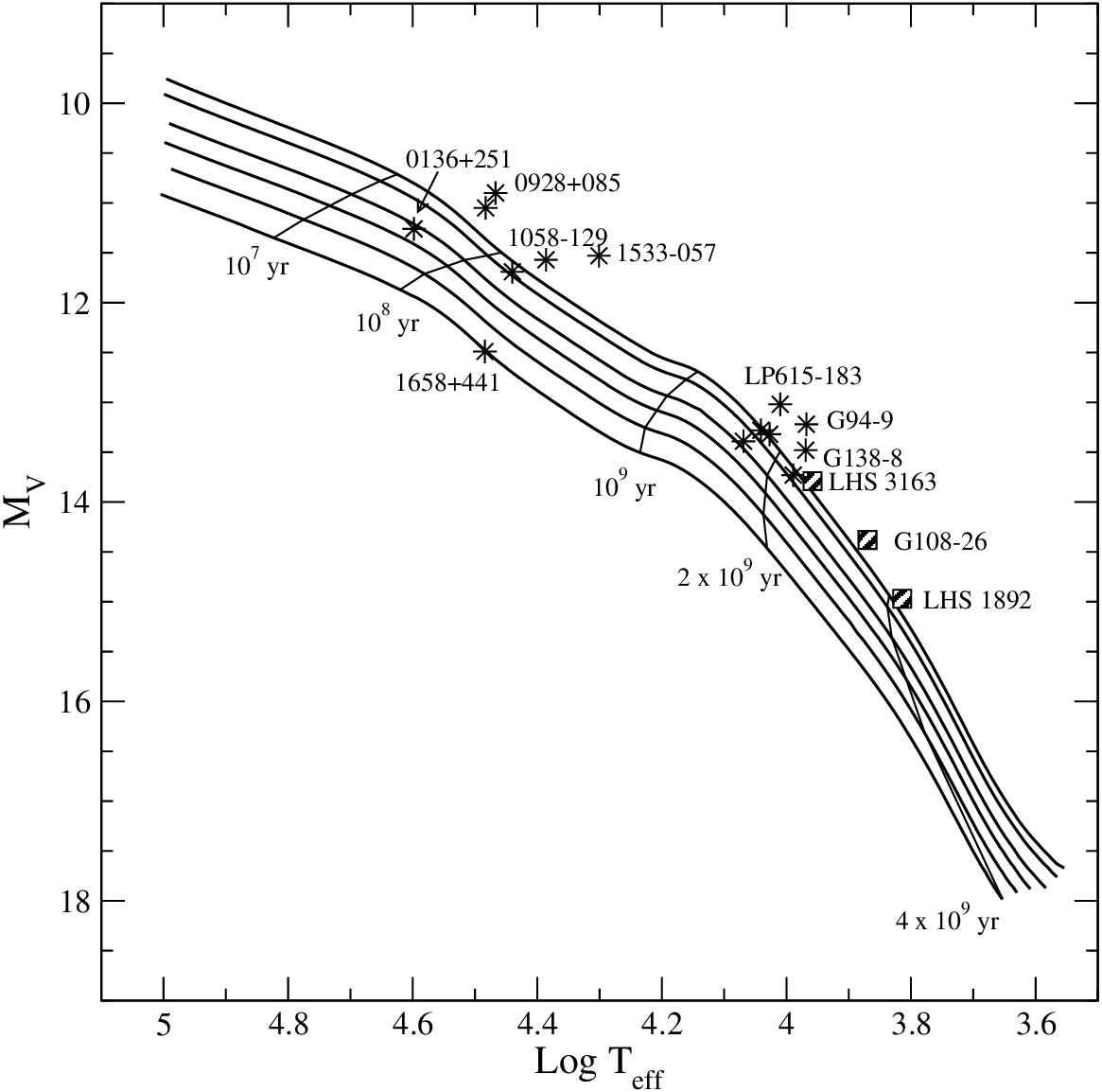}
\caption{Absolute  visual  magnitude  as  a  function of the effective 
         temperature  for  our  massive  pure H  white  dwarf  cooling
         sequences  with  masses of (from  top  to bottom) 1.06, 1.10, 
         1.16, 1.20, 1.24,  and $1.28\,M_{\sun}$.  The thin  lines plot  
         isochrones  at  the labelled  ages.  Observational  data  for 
         some  massive  H--rich  white  dwarfs from  Bergeron  et  al.  
         (2001) and  Liebert  et al.   (2005) are included --- squares 
         and stars, respectively.}
\label{mv_teff}
\end{figure}

In  Fig.~\ref{mv_teff} we illustrate  the run  of the  absolute visual
magnitude as a  function of the effective temperature  for our massive
pure H  white dwarf sequences.  In  addition, we draw 
various isochrones at ages of 0.01, 0.1,  1, 2, and 4 Gyr.  It is clear
that the  coolest white dwarf  models are not necessarily  the oldest.
For instance, in 4 Gyr the $1.28\,M_{\sun}$ models have cooled down to
4600 K, while the $1.06\,M_{\sun}$ model sequence remains much hotter
(7000 K)  at this  age.  Observational data  for massive  white dwarfs
with H--rich atmospheres taken from Bergeron et al. (2001) and Liebert
et  al. (2005)  are also  included in  the figure.  As  compared  with  
the ages  quoted  by  the
mentioned  authors, our  results suggest younger  ages  for the
white dwarfs  in Fig.~\ref{mv_teff}.   For instance, for  WD 1658+441,
which,  according  to  our  calculations  would  have  an  appreciable
fraction of its core in a crystallized state, we derive an age of 0.28
Gyr,   as  compared   with  the   0.38  Gyr   quoted  by   Liebert  et
el. (2005). The shorter ages  are due partly to the abundant $^{20}$Ne
in the core of our massive models, which results in a lower specific
heat per gram.

\begin{figure}[t]
\centering
\includegraphics[clip,width=250pt]{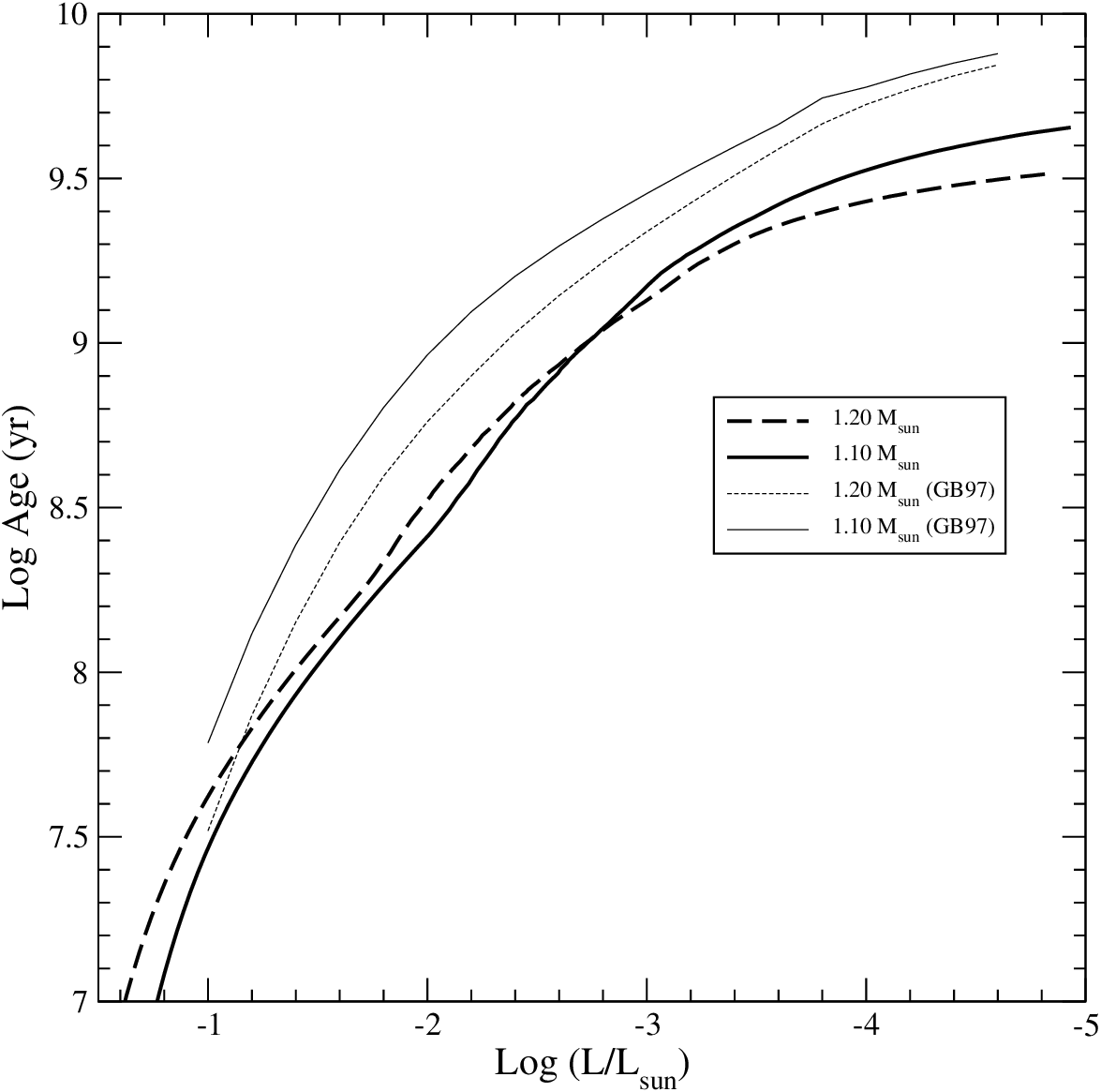}
\caption{Comparison between our $1.10$  and $1.20\,M_{\sun}$ sequences 
        (helium  atmospheres)  with  those  of  Garc\'{\i}a--Berro  et
        al. (1997a).}
\label{comparacion}
\end{figure}

Finally,   in  Fig.~\ref{comparacion}  we   compare  our   $1.10$  and
$1.20\,M_{\sun}$  sequences  with  helium  atmospheres with  those  of
Garc\'{\i}a--Berro  et al.   (1997a)  for ONe  cores.   Note that  our
calculations  predict  much  younger ages  than   those  derived  in
Garc\'{\i}a--Berro et  al. (1997a).  In part,  the differences between
both sets of calculations have  their origin in the different chemical
abundance  distributions  characterizing  the  pertinent  models.   For
instance, our  models are characterized  by a helium--dominated shell
that   is  four   times  more   massive  than  considered  in
Garc\'{\i}a--Berro et  al. (1997a).  It  should also be taken
into   account  that  the   preliminary  calculations   of
Garc\'{\i}a--Berro et  al. (1997a) were done using  a simplified model
that obtain the  evolution from  calculating  the
binding energy  very accurately (actually  using the same  equation of
state employed here)  and coupling it with a  relationship linking the
surface luminosity  with the core temperature of  an otherwise typical
CO white dwarf of mass $0.6\,M_{\sun}$.

In closing, we  list the main  characteristics of our 1.06,
1.16, and $1.28\,M_{\sun}$ white dwarf sequences in Table 1. Specifically, 
we list the effective temperature, the  surface gravity, the age, the
colors, and the  absolute visual magnitude.

\section{Conclusions}

We  have computed  the evolution  of  massive white  dwarf stars  with
oxygen--neon cores  for masses ranging from  1.06 to $1.28\,M_{\sun}$,
which covers the expected range of masses for which these stars should
presumably exist.  The evolutionary tracks for both pure H and pure He
envelopes  were computed for  surface luminosities  spanning the
range from  $\log(L/L_{\sun})\approx 0.5$ down to $-5.2$.   The use of
detailed non--gray  model atmospheres provides us  with accurate outer
boundary  conditions   for  our  evolving  models   at  low  effective
temperatures. To our  knowledge this is the first  attempt to compute
the evolution  of massive  white dwarfs with  a realistic  equation of
state --- which includes all  the non--ideal, corrective terms, and the
full temperature dependence --- and reliable chemical profiles for the
degenerate interior expected from the previous evolutionary history of
massive white  dwarf progenitors  that burned carbon  in semidegenerate
conditions.  We have examined  the cooling ages, colors, and magnitudes
of our sequences and find  that massive white dwarfs are characterized
by  very rapid evolution. Indeed, at  still observable luminosities,
we find that the cooling  of massive white dwarfs is largely dominated
by  Debye cooling with  the result  that these  white dwarfs  could be
unobservable  at ages  below the  age of  the Galactic  disk  with the
current observational  facilities.  At  such advanced stages,  we find
our sequences to have reached the turn--off in their colors and thus 
become blue in short times.

The results  presented here will  be helpful in  interpreting recent 
observations  of white dwarfs with very  high surface gravities
(Dahn et al.  2004; Madej et al.  2004; Nalezyty \& Madej 2004), which
up  to  now  rely  on  previous evolutionary  sequences that  were
computed  assuming carbon--oxygen  cores.  Finally,  we  prepared
detailed tabulations of ages, colors,  and magnitudes for all our white
dwarf  sequences,   which  are  available   at  our  web   site:  {\tt
http://www.fcaglp.unlp.edu.ar/evolgroup/}.

\begin{acknowledgements}
We acknowledge the valuable report of our referee, G.S.  Stringfellow,
which  strongly   improved  both  the  scientific   content  and 
presentation of  the paper. This  research was  supported by
the  MCYT grant AYA05--08013--C03--01  and 02,  by the  European Union
FEDER funds, by the AGAUR and by the PIP 6521 grant from CONICET.
\end{acknowledgements}

\end{document}